\documentclass[a4paper,showkeys,superscriptaddress,preprint,nofootinbib]{revtex4-1}
\usepackage[utf8]{inputenc}
\usepackage{textcomp}
\usepackage{amsmath}
\usepackage{amssymb}
\usepackage{graphicx}

\makeatletter

\pdfpageheight\paperheight
\pdfpagewidth\paperwidth

\@ifundefined{textcolor}{}
{%
 \definecolor{BLACK}{gray}{0}
 \definecolor{WHITE}{gray}{1}
 \definecolor{RED}{rgb}{1,0,0}
 \definecolor{GREEN}{rgb}{0,1,0}
 \definecolor{BLUE}{rgb}{0,0,1}
 \definecolor{CYAN}{cmyk}{1,0,0,0}
 \definecolor{MAGENTA}{cmyk}{0,1,0,0}
 \definecolor{YELLOW}{cmyk}{0,0,1,0}
}

\usepackage{bm}
\usepackage{indentfirst}
\usepackage{float}
\usepackage{subcaption}
\usepackage{caption}
\usepackage{psfrag}
\usepackage{epstopdf}
\usepackage{diagbox}
\usepackage{orcidlink}

\makeatother

\begin{document}
\title{Constraining tachyonic inflationary $\beta$-exponential model with
Continuous Spontaneous Localization collapse scheme}
\author{F. A. Brito \orcidlink{0000-0001-9465-6868}}
\email{fabrito@df.ufcg.edu.br}

\affiliation{Departamento de Física, Universidade Federal de Campina Grande, Caixa
Postal 10071, 58429-900 Campina Grande, Paraíba, Brazil}
\affiliation{Departamento de Física, Universidade Federal da Paraíba, Caixa Postal
5008, 58051-970 João Pessoa, Paraíba, Brazil}

\author{Julio C. M. Rocha \orcidlink{0000-0003-4515-9245}}
\email{julio.rocha@servidor.uepb.edu.br}
\affiliation{Centro de Ciências, Tecnologia e Saúde, Universidade Estadual da Paraíba, 58233-000, Araruna, PB, Brazil. }

\author{A. S. Lemos \orcidlink{0000-0002-3940-0779}}
\email{adiel@ufersa.edu.br}
\affiliation{Departamento de Ciências Exatas e Tecnologia da Informação, Universidade Federal Rural do Semi-Árido, 59515-000 Angicos, Rio Grande do Norte, Brazil}
\affiliation{Departamento de Física, Universidade Federal de Campina Grande, Caixa
Postal 10071, 58429-900 Campina Grande, Paraíba, Brazil}

\author{A. S. Pereira \orcidlink{0000-0002-5931-6224}}
\email{alfaspereira@gmail.com}

\affiliation{Departamento de Física, Universidade Federal de Campina Grande, Caixa
Postal 10071, 58429-900 Campina Grande, Paraíba, Brazil}
\affiliation{Instituto Federal da Para\'{i}ba, 58755-000 Princesa Isabel, Para\'{i}ba, Brazil}
\begin{abstract}
In this work, we consider the dynamics of the self-induced collapse
of the tachyon wave function in inflationary scenarios. We analyze
the modifications on the power spectrum by considering the $\beta$-exponential potential, whose parameters have updated constraints by the Planck 2018 baseline data and recent results from the Atacama Cosmology Telescope (ACT). Moreover, we show that for this kind of potential, just for a narrow range of $\beta$-parameter, there is agreement
between the theoretical predictions and the current observational
data. Considering the proposal for a collapse scheme that leads to
the modification of Schr\"odinger evolution of the inflation wave function
from the employment of a Continuous Spontaneous Localization (CSL)
approach, we derive the scalar spectral index and tensor-to-scalar
ratio. We then obtained the constraints on both collapse and $\beta$-parameters
that, in turn, yield deviations in the $n_{s}$ vs. $r$ plane when
compared to the $\beta$-exponential potential standard estimate.
The CSL scheme applied to tachyonic inflation driven by a $\beta$-potential
offers an adequate description of the recent data. 
\end{abstract}
\maketitle
\section{Introduction}

Cosmological observations strongly suggest that the early Universe
went through a phase of accelerated expansion known as inflation \citep{baumann2012tasilecturesinflation}.
The paradigm of inflation of the Universe has been investigated from
several scalar field theoretical models, whose nature is still undetermined
\citep{baumann2012tasilecturesinflation,arXiv:1303.3787}. In this context, we can highlight the
tachyon field, which was initially considered from studies by Sen
\citep{Sen:1998sm} about type II string theory and the tachyon instability
signals on D-branes. The cosmological consequences of the gravity-tachyon
system have been extensively researched (see \citep{Singh_2020,Gibbons_2003,FAIRBAIRN20021,FROLOV20028,Kofman_2002,PhysRevD.66.043530,Shiu:2002qe,PhysRevD.66.081301}
and related sources). For instance, Ref. \citep{Gibbons:2002md} studied
the cosmological relevance of the tachyon field for the expansion
of the Universe by considering several initial conditions.

Given the absence of compelling statistical evidence supporting a
particular inflationary model, a required current purpose is to investigate
the theoretical predictions of several classes of inflation models
in the context of present observational data \citep{baumann2012tasilecturesinflation,arXiv:1303.3787}.
In this work, we intend to examine theoretical estimates by studying
the $\beta$-exponential inflationary model (see Ref. \citep{Alcaniz_2007,Santos_2018,Santos_2022})
and relate them with the observational data. In turn, according
to the leading inflationary paradigm, in the inflationary era, the
evolution of the Universe is described by a Friedmann-Robertson-Walker
(FRW) background cosmology with an accelerated expansion driven by
the potential of a scalar field. Furthermore, the quantum fluctuations
of inflation are characterized by a initial vacuum state perfectly
homogeneous and isotropic \citep{baumann2012tasilecturesinflation,arXiv:1303.3787}. However, when
considering the transition from this state to the present non-symmetric
state of current Universe, we come across the fact that this transition
is not unitary, in other words, we have the measurement problem \citep{PhysRevD.94.043502}.

This scenario has been extensively discussed \citep{Perez_2006,doi:10.1142/S0218271811018937},
whose proposed solution is developed from the self-induced collapse
hypothesis, i.e., one considers a specific scheme by which a self
induced collapse of the wave function is taken as the mechanism by
which inhomogeneities and anisotropies arise at each particular scale.
In the present work, we will focus on the Continuous Spontaneous Localization
(CSL) model \citep{PhysRevLett.124.080402,Martin2021,Martin:2021lje}. In this case, we establish a formalism, known as the
CSL approach (see Refs. \citep{Leon_2021, GLEON} and references therein), for the slow-roll inflationary scenario in the context
of tachyonic inflation, aiming, thus, to obtain the spectra of scalar and tensor perturbations. Then, using observational data from Planck \citep{PLANCK,PLANCK2018}, and  from the Atacama Cosmology Telescope (ACT) Data Release 6 (DR6) \citep{calabrese2025atacamacosmologytelescopedr6},  we find constraints for the parameters of the $\beta$-exponential potential and CSL theoretical model. The recent ACT results refine the Planck measurements, achieving sensitivity comparable to that of the Planck legacy dataset, particularly in verifying the near-scale-invariance of primordial scalar perturbations.

This paper is organized as follows. In Sec. II, we briefly review
the tachyon inflation and discuss its features, specifically applied
to the $\beta$-potential model, as well as the quantum treatment
with the semiclassical gravity approximation. From the Planck $2018$
baseline analysis, in addition to BK18 \& BAO data and ACT DR6 data, we get constraining the $\beta$-parameter. In Sec. III, we discuss the CSL scheme and apply it to a tachyonic inflationary model to obtain the scalar power spectrum for all modes. In Sec. IV, we present the results of our analysis and find constraints on the collapse- and $\beta$-parameters by analyzing the current observational data for the scalar spectral index as well as the tensor-to-scalar ratio. Finally, in Sec. V, we summarize the main results and present the final remarks. Throughout
this work we use $\hbar=1$ and the metric signature is $\left(-,+,+,+\right)$.

\section{Tachyon Inflation}

In this section, we study the model of tachyonic inflation starting
from an effective action of the Dirac-Born-Infeld (DBI) type, which
includes the tachyon \citep{GAROUSI,2018MNRAS.481.2393F,Rasouli2019,Karami_2013,KARAMI2010216}. The model under consideration
is described by a $4D$ effective action of tachyon coupled to gravity
as follows \citep{LUNIN,REZAZADEH2020168299}, 
\begin{equation}
S=\int{d^{4}x\,\sqrt{-g}\left(\dfrac{1}{2\kappa^{2}}R-V(T)\sqrt{1+\alpha^{\prime}\partial_{\mu}T\partial^{\mu}T}\right)},\label{action}
\end{equation}
where $\kappa^{2}=1/M_{p}^{2}=8\pi G$ sets the 4D Planck scale, $M_{P}=1.2\times10^{19}\mathrm{GeV/c^{2}}$
is the Planck mass, $\alpha^{\prime}=l_{s}^{2}$ with $l_{s}$ being
the string scale. As previously stated,
in this work, we will consider the $\beta$-exponential potential
\citep{Santos_2018}. For the usual case, $\beta\rightarrow0$, the potential $V\left(T\right)$ is a function of
the tachyon real scalar field $T$, which satisfies conditions $V\left(0\right)<\infty$,
$V_{T}\left(T>0\right)<0$ and $V\left(|T|\rightarrow\infty\right)\rightarrow0$, with the
notation $V_{T}=dV/dT$ \citep{STEER2004, PhysRevD.95.103506}.

Now, for a spatially homogeneous metric, i.e., $g_{\mu\nu}=\mbox{diag}(-1,a^{2}(t),a^{2}(t),a^{2}(t))$,
the Einstein field equations, $G_{\nu}^{\mu}=8\pi GT_{\nu}^{\mu}$,
arrive in Friedmann equations, where 
\begin{equation}
T_{\nu}^{\mu}=V(T)\left[\dfrac{\alpha^{\prime}g^{\mu\lambda}\partial_{\nu}T\partial_{\lambda}T-\delta_{\nu}^{\mu}(1+\alpha^{\prime}\partial_{\lambda}T\partial^{\lambda}T)}{\sqrt{1+\alpha^{\prime}\partial_{\lambda}T\partial^{\lambda}T}}\right]\label{stress}
\end{equation}
is the energy-momentum tensor for the tachyon field, which we shall
take to be of the general perfect fluid form, $T_{\nu}^{\mu}=\mbox{diag}\left(-\rho,p,p,p\right)$.
In this case, the dynamics is governed by Friedmann equations 
\begin{equation}
H^{2}=\left(\frac{\dot{a}}{a}\right)^{2}=\frac{\rho}{3M_{p}^{2}},\quad\dot{H}=-\frac{1}{2M_{p}^{2}}(\rho+p),\label{eq:friedmanneq}
\end{equation}
where the overdot represent derivative with respect to time, $\dot{H}=dH/dt$,
and the energy density and pressure are given by 
\begin{equation}
\rho=\frac{V({T})}{\sqrt{1-\alpha^{\prime}\dot{T}^{2}}},\quad p=-V(T)\sqrt{1-\alpha^{\prime}\dot{T}^{2}}.
\end{equation}
In turn, the equation of state parameter for the tachyon scalar
field can then be written as 
\begin{equation}
w=\frac{p}{\rho}=\alpha^{\prime}\dot{T}^{2}-1=-c_{s}^{2},\label{eq:eqofstate}
\end{equation}
where $c_{s}$ is the effective sound speed \citep{PIAZZA, 2018ApJ...853..188A}. The dynamics
of the tachyon scalar field are governed by equation 
\begin{equation}
\frac{1}{\sqrt{-g}}\partial_{\mu}\left(\sqrt{-g}g^{\mu\nu}\partial_{\nu}T\right)-\frac{\partial_{\mu}(\alpha^{\prime}\partial_{\lambda}T\partial^{\lambda}T)\partial^{\mu}T}{2(1+\alpha^{\prime}\partial_{\lambda}T\partial^{\lambda}T)}-\frac{V_{T}}{\alpha^{\prime}V}=0.
\end{equation}
In case of a spatially homogenous tachyon field in FRW spacetime,
the equation of motion reduces to 
\begin{equation}
\dfrac{\ddot{T}}{1-\alpha^{\prime}\dot{T}^{2}}+3H\dot{T}+\frac{V_{T}}{\alpha^{\prime}V}=0\,\mbox{.}\label{eq:scalar_dynamics}
\end{equation}

In context of the slow-roll inflationary regime, $\alpha^{\prime}\dot{T}^{2}\ll1$
and $|\ddot{T}|\ll3H|\dot{T}|$, we find 
\begin{equation}
H^{2}\approx\dfrac{V}{3M_{p}^{2}}\mbox{,}\quad\dot{H}\approx-\frac{\alpha^{\prime}V\dot{T}^{2}}{2M_{p}^{2}}\mbox{,}\quad3H\dot{T}+\frac{V_{T}}{\alpha^{\prime}V}\approx0.
\end{equation}
Now we can to define slow-roll parameters for tachyon inflation, i.e.,
\begin{eqnarray}
\varepsilon_{1}\, & \equiv & \,-\dfrac{\dot{H}}{H^{2}},\nonumber \\
\varepsilon_{i+1}\, & \equiv & \,\dfrac{\dot{\varepsilon_{i}}}{H\varepsilon_{i}}.
\end{eqnarray}
In turn, the first slow-roll parameters are related to the inflation
potential as follows 
\begin{eqnarray}
\varepsilon_{1}\, & = & \,\dfrac{M_{p}^{2}}{2\alpha^{\prime}}\frac{V_{T}^{2}}{V^{3}},\\
\varepsilon_{2}\, & = & \,\dfrac{M_{p}^{2}}{\alpha^{\prime}}\left(-\frac{2V_{TT}}{V^{2}}+\frac{3V_{T}^{2}}{V^{3}}\right).
\end{eqnarray}
Note that both, $\varepsilon_{1}$ and $\varepsilon_{2}$, can have
either sign. However, the slow-roll conditions impose that $\varepsilon_{1}\ll3/2$,
and $\varepsilon_{2}\ll6$.

If we start the field at a value $T_{k}$, the number of e-folds before
the slow-roll parameters becomes of order unity (that is, before inflation
ends, $T_{e}$) can be written in terms of the tachyonic potential.
Hence, the number of e-folds of the inflation produced when the tachyon
field rolls from a particular value $T_{k}$ to end point $T_{e}$
is 
\begin{equation}
N(T_{k},T_{e})=\int_{t}^{t_{e}}H(t)dt=\int_{T_{k}}^{T_{e}}\dfrac{H}{\dot{T}}dT=\dfrac{\alpha^{\prime}}{M_{p}^{2}}\int_{T_{e}}^{T_{k}}\dfrac{V^{2}}{V_{T}}dT.\label{Number}
\end{equation}

In order to solve the horizon problem, generally, it is required that
the accelerated period be supported for $60$ or more e-folds. In
the following subsection, we must apply these results by considering
the $\beta$-exponential potential, aiming to find constraints for
this theoretical model from Planck data \citep{PLANCK,PLANCK2018}.

\subsection{$\beta$-exponential potential}

The $\beta$-exponential potential, initially phenomenologically proposed
\citep{Alcaniz_2007} and after recovered in the context of braneworld
cosmology \citep{Santos_2018}, is a class of potentials employed as a
generalization of the usual inflationary exponential potential. In
this scenario, some studies have determined cosmological solutions
for a wide range of $\beta$-values \citep{arXiv:1303.3787}. Hence, in
the context of tachyonic inflation, we must consider the $\beta$-exponential
potential: 
\begin{equation}
V\left(T\right)=V_{0}\left(1-\beta\lambda T\right)^{1/\beta},\label{betaexponential}
\end{equation}
where $\beta$ is a free parameter to be constrained by the appropriate
choice of values that yield the model predictions compatible with
the empirical data.

\begin{figure}[th]
\includegraphics[scale=0.3]{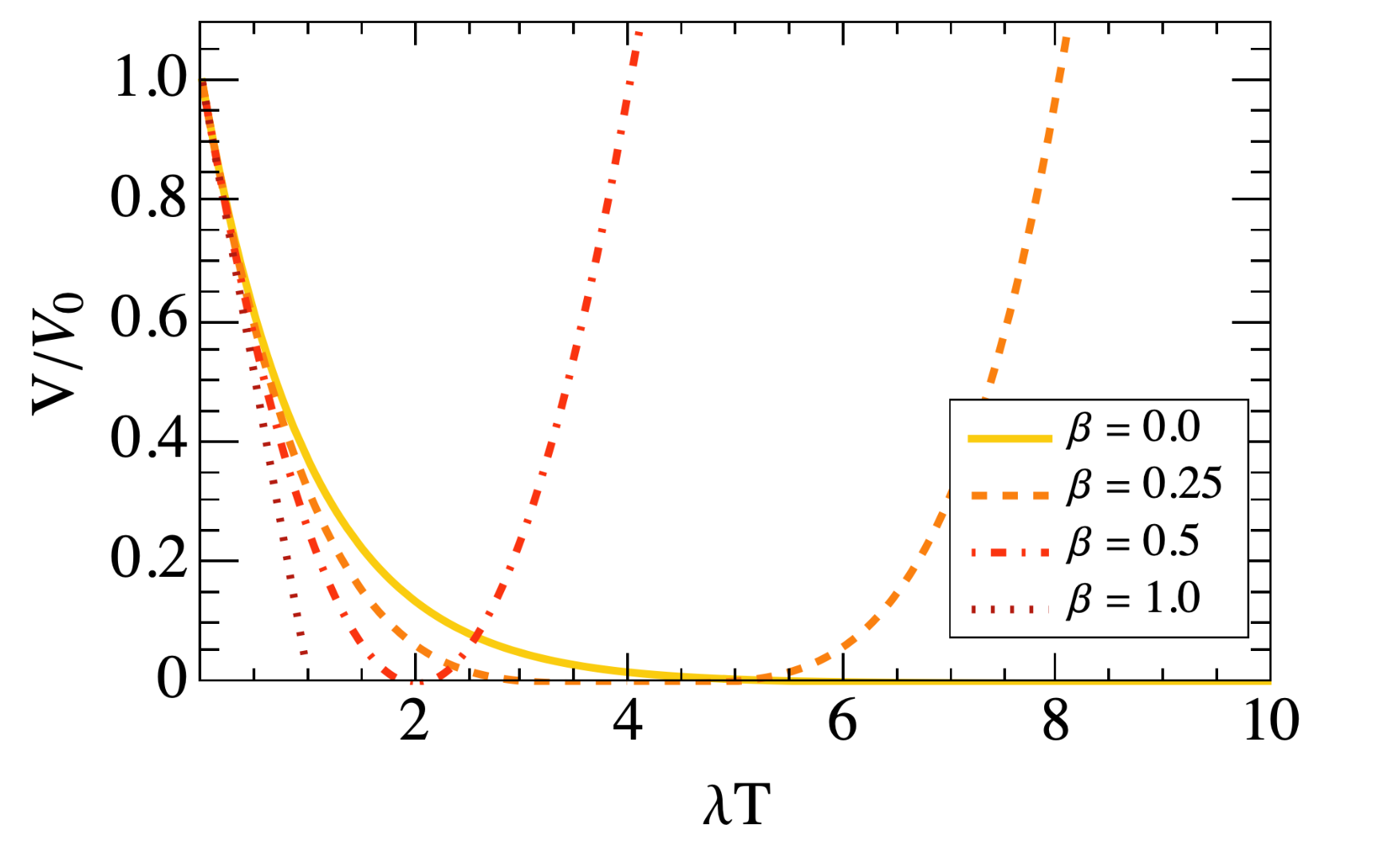}\caption{The potential $V(T)$ as a function of the tachyon field $T$ is shown
to some $\beta$-values. }

\label{fig1} 
\end{figure}

Since $\lim_{\beta\rightarrow0}V(T)=\exp(-\lambda T)$, Eq. (\ref{betaexponential})
can be seen as a generalization of the usual inflationary exponential
potential. In Fig. \ref{fig1}, the behavior of the $\beta$-exponential
potential as a function of the field $T$ is shown for increasing
values of the $\beta$-parameter. 

For the potential (\ref{betaexponential}), the slow-roll parameters
is 
\begin{equation}
\varepsilon_{1}(T)=\dfrac{M_{p}^{2}\lambda^{2}}{2\alpha^{\prime}V_{0}}\left(1-\beta\lambda T\right)^{-2-1/\beta},\qquad\varepsilon_{2}(T)=\dfrac{M_{p}^{2}\lambda^{2}(1-2\beta)}{\alpha^{\prime}V_{0}}\left(1-\beta\lambda T\right)^{-2-1/\beta},
\end{equation}
while the scalar spectral index and the tensor-to-scalar ratio are
given by 
\begin{equation}
n_{s}=1-\dfrac{2M_{p}^{2}\lambda^{2}}{\alpha^{\prime}V_{0}}(1+\beta)\left(1-\beta\lambda T\right)^{-2-1/\beta},\quad r=\dfrac{8M_{p}^{2}\lambda^{2}}{\alpha^{\prime}V_{0}}\left(1-\beta\lambda T\right)^{-2-1/\beta}.\label{spectralindex}
\end{equation}
On the other hand, through Eq. (\ref{Number}), it is straightforward
to obtain the number of e-folds, 
\begin{equation}
N=\dfrac{\alpha^{\prime}V_{0}}{M_{p}^{2}\lambda^{2}(1+2\beta)}\left[\left(1-\beta\lambda T_{k}\right)^{2+1/\beta}-\left(1-\beta\lambda T_{e}\right)^{2+1/\beta}\right],
\end{equation}
where $T_{e}$ is the value of $T$ when inflation ends, i.e., $\varepsilon_{1}(T_{e})=1\Rightarrow\left(1-\beta\lambda T_{e}\right)^{2+1/\beta}=M_{p}^{2}\lambda^{2}/2\alpha^{\prime}V_{0}$,
then 
\begin{equation}
\left(1-\beta\lambda T_{k}\right)^{2+1/\beta}=\dfrac{M_{p}^{2}\lambda^{2}}{2\alpha^{\prime}V_{0}}\left[1+2N(1+2\beta)\right].\label{tn}
\end{equation}
Therefore, we can rewrite the slow-roll parameters as a function of
$N$ as follows: 
\begin{equation}
\varepsilon_{1}=\dfrac{1}{1+2N(1+2\beta)},\quad\varepsilon_{2}=\dfrac{2+4\beta}{1+2N(1+2\beta)}.
\end{equation}
Now, using (\ref{spectralindex}), we get 
\begin{equation}
n_{s}=1-\dfrac{4(1+\beta)}{1+2N(1+2\beta)},\quad r=\dfrac{16}{1+2N(1+2\beta)}.\label{nindex}
\end{equation}

\begin{figure}[t]
\includegraphics[scale=0.36]{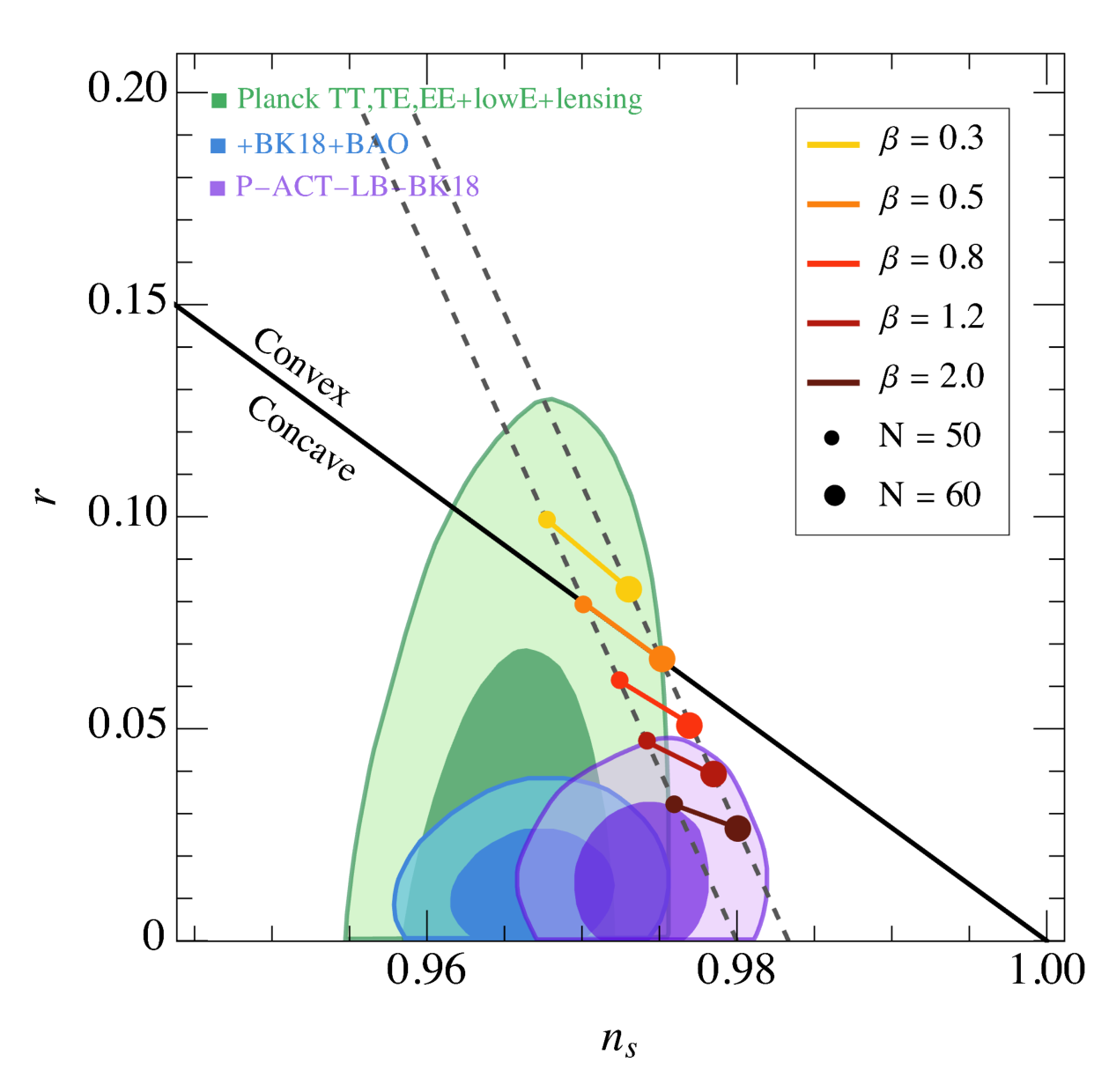} \caption{The marginalized joint regions at $68\%$ and $95\%$ confidence levels
for the spectral index ($n_{s}$) and tensor-to-scalar ratio ($r$)
derived from Planck data (independently and combined with BK18+BAO) \citep{PLANCK2018} and Atacama Cosmology Telescope (combined with other datasets) \citep{calabrese2025atacamacosmologytelescopedr6} are compared to the theoretical predictions
of the tachyonic inflationary $\beta$-exponential model. The number
of e-folds have been fixed at $N=50$ and $N=60$.}
\label{fig2} 
\end{figure}

The $r$ vs. $n_{s}$ plane shown in Fig. \ref{fig2} presents constraints
derived from the \textit{Planck 2018} baseline analysis. Additionally,
incorporating BK18 \& BAO \citep{PLANCK2018} and ACT DR6 \citep{calabrese2025atacamacosmologytelescopedr6}, one refines and narrows
these constraints, which improves the bounds on the primordial gravitational
waves parameterized by the tensor-to-scalar ratio $r$. Besides, the
inclined thick line divides the $r$ vs. $n_{s}$ plane between convex
and concave potentials. So, for values of $\beta<0.8$, the $\beta$-exponential
potential behaves as a convex potential otherwise presents a concave
shape. Moreover, at $95\%$ CL, from Planck TT,TE,EE+lowE+lensing
likelihood, the tachyonic inflation is excluded for $\beta\geq2.0$
($\beta\leq2\times10^{-2}$) with $N\geq50$ ($N\leq60$). On the
other hand, in light of the joint Planck Collaboration $2018$ baseline
analysis, when adding BK18 \& BAO, the $\beta$-exponential tachyonic
inflation model is ruled out for any $\beta$-values. Conversely, incorporating the ACT dataset reveals that parameter values of $\beta\geq1.2$ are consistent with the observational bounds while providing sufficient inflation. Finally, as
$\beta$-parameter increases, the tachyonic inflation prediction converges for the $\beta$-exponential standard estimate \citep{Santos_2018}.

In the next subsection, we discuss the semiclassical treatment for
this theoretical model aiming to obtain the relation between quantum
and classical measurements.

\subsection{Classical description of the perturbations}

In proceeding to analyze the perturbations, we choose to work in the longitudinal
gauge, and now focusing on the scalar perturbations at first order. For this case, the line element associated to the metric is
\begin{equation}
ds^{2}=-(1+2\Phi)dt^{2}+a^{2}(t)(1-2\Phi)\delta_{ij}dx^{i}dx^{j},\label{longitudinal}
\end{equation}
where $\Phi$ represents the scalar perturbation, which in the Newtonian
limit is identified with the effective gravitational potential. Using
equation (\ref{stress}) with the metric element of longitudinal gauge
from equation (\ref{longitudinal}), we find that the perturbed Einstein's
equation $\delta G_{\nu}^{\mu}=8\pi G\delta T_{\nu}^{\mu}$ for tachyon
scalar field can be calculated, in Fourier modes, and provides us
\citep{Singh_2020}: 
\begin{align}
3H^{2}\Phi+3H\dot{\Phi}+\frac{k^{2}\Phi}{a^{2}}=-4\pi G\delta\rho,\label{eq:Eineq_1}\\
\ddot{\Phi}+4H\dot{\Phi}+\left(2\dot{H}+3H^{2}\right)\Phi=4\pi G\delta p,\label{eq:Eineq_2}\\
\dot{\Phi}+H\Phi=\dfrac{3H^{2}}{2}\alpha^{\prime}\dot{\bar{T}}\delta T,\label{eq:Eineq_3}
\end{align}
where 
\begin{align}
\delta\rho & =\dfrac{\dot{H}}{4\pi G}\frac{\Phi}{\left(1-\alpha^{\prime}\dot{\bar{T}}^{2}\right)}+\dfrac{3H^{2}}{8\pi G}\frac{\alpha^{\prime}\dot{\bar{T}}\delta\dot{T}}{\left(1-\alpha^{\prime}\dot{\bar{T}}^{2}\right)}+\dfrac{V_{T}\delta T}{\sqrt{1-\alpha^{\prime}\dot{\bar{T}}^{2}}},\\
\delta p & =\dfrac{\dot{H}}{4\pi G}\Phi+\dfrac{3H^{2}}{8\pi G}\alpha^{\prime}\dot{\bar{T}}\delta\dot{T}-V_{T}\delta T\sqrt{1-\alpha^{\prime}\dot{\bar{T}}^{2}}.
\end{align}

The curvature perturbation $\zeta$ on the uniform field slicing is defined as a gauge-invariant combination of scalar field
 perturbation $\delta T$ and the metric perturbation $\Phi$ \cite{2024PDU....4601560Y}:
 \begin{equation}
     \zeta = \Phi + \left(\frac{H}{\dot{\bar{T}}}\right)\delta T.
 \end{equation}
 The Mukhanov-Sasaki variable is
 \begin{equation}
        v = z\zeta,
\end{equation}
 where 
 \begin{equation}
     z = \frac{a(\bar{\rho} + \bar{p})^{1/2}}{c_s H} = \frac{\sqrt{3\alpha^{\prime}}aM_{p}\dot{\bar{T}}}{c_s},
 \end{equation}
 and therefore
 \begin{equation}
     v = z\left[\Phi + \left(\frac{H}{\dot{\bar{T}}}\right)\delta T\right]. 
\end{equation}

Combining Eqs. (\ref{eq:Eineq_1}) and (\ref{eq:Eineq_3}) and performing a change of variables of
the time $t$ by the conformal time, $dt=a(\eta)d\eta$, we obtain
 \begin{equation}
  \nabla^{2}\Phi - \frac{\mathcal{H}^{\prime}}{a}\Phi = \sqrt{\frac{\varepsilon_1}{2}}\frac{H}{a^{2}M_{p}c_{s}^{2}}\left(v^{\prime} - \frac{z^{\prime}}{z}v\right).
\end{equation}
 In Fourier modes, considering $k^{2}\gg aH$, we get
 \begin{equation}
 \Phi_{k} = \sqrt{\frac{\varepsilon_1}{2}}\frac{H}{a^{2}c_{s}^{2}k^{2}M_{p}}\left(v_{k}^{\prime} - \frac{z^{\prime}}{z}v_{k}\right).
 \end{equation}
 This way leads to quantization of $\Phi$ through quantization of the Mukhanov-Sasaki variable.

\section{Collapse Spontaneous Localization (CSL)}

The self-induced collapse hypothesis of the inflaton wave function
has been discussed as a likely physical process which leads to the
emergence of inhomogeneity and anisotropy \citep{LEON1,PICCIRILLI,CANATE,GLEON}.
In this section, we aim to start the treatment of the quantum theory
for the field $\delta\hat{T}$. For this purpose, one can 
expand the action (\ref{action}) up to second order in the Mukhanov-Sasaki variable $v$, when finds
\begin{equation}
S=\dfrac{1}{2}\int d\eta\, d^{3}\mathbf{k}\left[v_{\mathbf{k}}^{\prime}v_{\mathbf{k}}^{\star\prime}-c_{s}^{2}k^{2}v_{\mathbf{k}}v_{\mathbf{k}}^{\star}-\dfrac{z^{\prime}}{z}\left(v_{\mathbf{k}}v_{\mathbf{k}}^{\star\prime}+v_{\mathbf{k}}^{\prime}v_{\mathbf{k}}^{\star}\right)+\dfrac{z^{\prime\,2}}{z^{2}}v_{\mathbf{k}}v_{\mathbf{k}}^{\star}\right].\label{act1}
\end{equation}
Here, both fields $v_{\mathbf{k}}$ and the canonical conjugated momentum $\pi_{\mathbf{k}}=v_{\mathbf{k}}^{\prime}-(\hat{z}^{\prime}/\hat{z})v_{\mathbf{k}}$
satisfy the commutation relations $[\hat{v}_{\mathbf{k}},\hat{\pi}_{\mathbf{k}^{\prime}}]=i\delta(\mathbf{k} - \mathbf{k}^{\prime})$.

The Collapse Spontaneous Localization (CSL) model is defined from
a non-unitary modification of the Schr\"odinger equation that induces a wave function collapse towards one of the possible eigenstates of an operator called the collapse operator \citep{PICCIRILLI}. In this sense, it is convenient to describe the theory in terms of the Hamiltonian
of the system in Fourier modes, which in this case is given by 
\begin{equation}
\hat{H}_{\mathbf{k}}^{R,I}=\dfrac{1}{2}\int d^{3}k\left[\hat{\pi}_{\mathbf{k}}^{R,I}\hat{\pi}_{\mathbf{k}}^{*R,I}+c_{s}^{2}k^{2}\hat{v}_{\mathbf{k}}^{R,I}\hat{v}_{\mathbf{k}}^{*R,I}-\dfrac{(1+\varepsilon_{1}+\varepsilon_{2}/2)}{\eta}\left(\hat{v}_{\mathbf{k}}^{R,I}\hat{\pi}_{\mathbf{k}}^{*R,I}+\hat{v}_{\mathbf{k}}^{*R,I}\hat{\pi}_{\mathbf{k}}^{R,I}\right)\right],
\end{equation}
where the indexes $R$,$I$ denote the real and imaginary parts of
$\hat{v}_{\mathbf{k}}$ and $\hat{\pi}_{\mathbf{k}}$, respectively. Defining $\Phi[\pi]$ the wave functional characterizing the quantum state of the field, the time evolution of this wave function is given by
\begin{equation}
    \left|\Phi,t\right\rangle =\hat{T}\exp\left\{ -\int_{t_{0}}^{t}dt^{\prime}\left[i\hat{H}+\frac{\left(\mathcal{W}\left(t^{\prime}\right)-2\lambda\hat{C}\right)^{2}}{4\lambda}\right]\right\} \left|\Phi,t_{0}\right\rangle. 
\end{equation}
Then, we can factorize $\Phi[\pi]$ into mode component $\Phi[\hat{v}_{\mathbf{k}}]=\Pi_{\mathbf{k}}\Phi[{v_{\mathbf{k}}^{R}}]\,\times\,\Phi[{v_{\mathbf{k}}^{I}}]$.
Here, we are aiming to deal with each mode separately. Since the Hamiltonian is quadratic in $\hat{v}_{\mathbf{k}}^{R,I}$ and $\hat{\pi}_{\mathbf{k}}^{R,I}$, is natural to assume a Gaussian state 
\begin{equation}
\Phi^{R,I}(\eta,\hat{\pi}_{\mathbf{k}}^{R,I})=\exp[-A_{k}(\eta)(\pi_{\mathbf{k}}^{R,I})^{2}+B_{k}(\eta)\pi_{\mathbf{k}}^{R,I}+C_{k}(\eta)],\label{Eq:eom1}
\end{equation}
with the wave function evolving according to Schr\"odinger equation,
and satisfying the initial conditions given by $A_{k}(\tau)=1/2k$,
$B_{k}(\tau)=C_{k}(\tau)=0$. So, the evolution of the state vector
characterizing the inflaton in Fourier modes as given by the CSL theory,  is assumed
to be 
\begin{equation}
\left|\Phi_{\mathbf{k}}^{R,I},\eta\right\rangle =\hat{\mathcal{T}}\exp\left\{ -\int_{\tau}^{\eta}d\eta^{\prime}\left[i\hat{H}_{\mathbf{k}}^{R,I}+\dfrac{(\mathcal{W}(\eta^{\prime})-2\lambda_{k}\hat{\pi}_{\mathbf{k}}^{R,I})^{2}}{4\lambda_{k}}\right]\right\} \left|\Phi_{\mathbf{k}}^{R,I},\tau\right\rangle ,
\end{equation}
where $\hat{\mathcal{T}}$ is the time-ordering operator and $\mathcal{W}\left(\eta'\right)$
is the background noise that can be considered as a stochastic process with continuous time. Using the solution (\ref{Eq:eom1}) and the CSL evolution equations, it can be shown that \citep{CANATE} 
\begin{equation}
\overline{\langle\hat{\pi}_{\mathbf{k}}^{R,I}\rangle^{2}}=\overline{\langle(\hat{\pi}_{\mathbf{k}}^{R,I})^{2}\rangle}-\dfrac{1}{4\mbox{Re}{A_{k}(\eta)}}.\label{media}
\end{equation}
Note that $\left(4\mbox{Re}{A_{k}(\eta)}\right)^{-1}$ represents
the variance of the momentum operator.

In this point, we can relate this quantity to the scalar power spectrum defined through the relation 
\begin{equation}
\overline{\Phi_{\mathbf{k}}\Phi_{\mathbf{k}^{\prime}}^{*}}=\dfrac{\varepsilon_{1}H^{2}}{2c_{s}^{4}k^{4}a^{4}M_{p}^{2}}\overline{\langle\hat{\pi}_{\mathbf{k}}\rangle\ \langle\hat{\pi}_{\mathbf{k}^{\prime}}\rangle^{*}}=\frac{2\pi^{2}}{k^{3}}\mathcal{P}_{s}(k)\delta({\mathbf{k}}-\mathbf{k}^{\prime}),\label{5.3.1}
\end{equation}
where $\mathcal{P}_{s}(k)$ is the scalar power spectrum \citep{HWANG}.

On the other hand, admitting that 
\begin{align}
\overline{\langle\hat{\pi}_{\mathbf{k}}\rangle\ \langle\hat{\pi}_{\mathbf{k}^{\prime}}\rangle^{*}} & =\overline{\langle\hat{\pi}_{\mathbf{k}}^{R}+i\hat{\pi}_{\mathbf{k}}^{I}\rangle\ \langle\hat{\pi}_{\mathbf{k}^{\prime}}^{R}-i\hat{\pi}_{\mathbf{k}^{\prime}}^{I}\rangle}\nonumber \\
 & =\left(\overline{\langle\hat{\pi}_{\mathbf{k}}^{R}\rangle^{2}}+\overline{\langle\hat{\pi}_{\mathbf{k}}^{I}\rangle^{2}}\right)\delta({\mathbf{k}}-\mathbf{k}^{\prime}),
\end{align}
we obtain 
\begin{equation}
\mathcal{P}_{s}(k) =  \dfrac{\varepsilon_{1}H^{2}}{4\pi^{2}c_{s}^{4}a^{4}M_{p}^{2}}\dfrac{\left(\overline{\langle\hat{\pi}_{\mathbf{k}}^{R}\rangle^{2}}+\overline{\langle\hat{\pi}_{\mathbf{k}}^{I}\rangle^{2}}\right)}{k}.\label{powerPhi}
\end{equation}
Furthermore, the expected value in (\ref{media}) is straightforward
obtained by assuming that wave function have the form (\ref{Eq:eom1}),
\begin{eqnarray}
\overline{\langle(\pi_{\mathbf{k}}^{R,I})^{2}\rangle} & \simeq & \frac{c_{s}k}{\pi}2^{2\nu_{s}-2}\Gamma^{2}(\nu_{s})\left[1+\lambda_{k}\sin\gamma_{k}\cos\gamma_{k}\right.\nonumber \\
 & - & \left.\frac{\lambda_{k}c_{s}k\tau}{2}\left(\frac{3}{\nu_{s}+1}\sin^{2}\gamma_{k}+\frac{\cos^{2}\gamma_{k}}{\nu_{s}}\right)\right](-c_{s}k\eta)^{-2\nu_{s}+1},\label{firstterm}
\end{eqnarray}
with $\gamma_{k}\equiv-c_{s}k\tau-\nu_{s}\pi/2-3\pi/4$. In turn,
the second term of r.h.s. in (\ref{media}) is given by 
\begin{equation}
\dfrac{1}{4\mbox{Re}{[A_{k}(\eta)]}}\simeq\frac{c_{s}k2^{2\nu_{s}-2}\zeta_{k}^{-2\nu_{s}}\sin(\pi\nu_{s})\Gamma^{2}(\nu_{s})(-c_{s}k\eta)^{-2\nu_{s}+1}}{\pi\sin(2\nu_{s}\theta_{k}+\pi\nu_{s})},\label{secondterm}
\end{equation}
in which $\zeta_{k}\equiv(1+4\lambda_{k}^{2})^{1/4},$ and $\theta_{k}\equiv-1/2\arctan(2\lambda_{k})$
\citep{GLEON}.

Finally, substituting Eqs. (\ref{media}), (\ref{firstterm}), and
(\ref{secondterm}) into (\ref{powerPhi}), we get 
\begin{eqnarray}
\mathcal{P}_{s}(k) &=& \dfrac{\varepsilon_{1}H^{2}}{2\pi^{2}c_{s}^{4}a^{4}M_{p}^{2}}2^{2\nu_{s}-2}\Gamma^{2}(\nu_{s})(-c_{s}k\eta)^{-2\nu_{s}+1}\mathcal{F}(\lambda_{k},\nu_{s})\nonumber \\ 
&=& \dfrac{\varepsilon_{1}H^{2}}{2\pi^{2}c_{s}^{4}a^{4}M_{p}^{2}}2^{2\nu_{s}-2}\Gamma^{2}(\nu_{s})(1-\varepsilon_{1})^{-2\nu_{s}+1}\left(\frac{c_s k}{aH}\right)^{-2\nu_{s}+1}\mathcal{F}(\lambda_{k},\nu_{s}),\label{power}
\end{eqnarray}
where $\mathcal{F}(\lambda_{k},\nu_{s})$ is a function of the collapse parameter, given by 
\begin{align}
\mathcal{F}(\lambda_{k},\nu_{s}) & =1+\lambda_{k}\sin\gamma_{k}\cos\gamma_{k}-\dfrac{\lambda_{k}c_{s}k\tau}{2}\left(\dfrac{3}{\nu_{s}+1}\sin^{2}\gamma_{k}+\dfrac{\cos^{2}\gamma_{k}}{\nu_{s}}\right)\nonumber \\
 & -\dfrac{\sin(\pi\nu_{s})}{\zeta_{k}^{2\nu_{s}}\sin(2\nu_{s}\theta_{k}+\pi\nu_{s})}.\label{F_lambda_k}
\end{align}

Now, from Eq. (\ref{power}), we can identify $\mathcal{P}_{s}(k)$
as follows: 
\begin{equation}
\mathcal{P}_{s}(k)=\mathcal{A}_{s}(k)\left(\frac{c_s k}{aH}\right)^{n_s - 1}\mathcal{F}(\lambda_{k},\nu_{s}),\label{scalarpwr}
\end{equation}
where $\mathcal{A}_{s}$ is the amplitude of the scalar power spectrum,
given by \citep{baumann2012tasilecturesinflation} 
\begin{equation}
\mathcal{A}_{s}=\frac{H^{2}}{8\pi^{2}M_{p}^{2}c_{s}\varepsilon_{1}}\,\mbox{.}
\end{equation}
On the other hand, for tensor perturbations, as shown in \citep{GLEON},
the tensor collapse function is analogous to the scalar collapse function,
and therefore 
\begin{equation}
\mathcal{P}_{t}(k)=\mathcal{A}_{t}(k)\left(\frac{c_s k}{aH}\right)^{n_s - 1}\mathcal{F}(\lambda_{k},\nu_{t}),\label{DUPLICATA: scalarpwr}
\end{equation}
where $\nu_{t}=1/2+\varepsilon_{1}$ and the amplitude of the tensor power
spectrum is given by \citep{baumann2012tasilecturesinflation} 
\begin{equation}
\mathcal{A}_{t}=\frac{2H^{2}}{\pi^{2}M_{p}^{2}}.
\end{equation}
Therefore, these results indicate that the collapse parameter $\lambda_{k}$
modifies the behavior of the power spectrum, and such deviations,
eventually, can be measured in experiments of the Cosmic Microwave
Background (CMB) spectrum, for instance. Furthermore, it has been
showed that when $\lambda_{k}=\lambda_{0}/c_{s}k$, the primordial
power spectrum becomes nearly scale-invariant \citep{Leon_2021,CANATE, PICCIRILLI}. This
leads to define 
\begin{equation}
\lambda_{k}\equiv\lambda_{0}\left(\dfrac{1}{c_{s}k}+\dfrac{\alpha}{c_{s}^{2}k^{2}}\right),\label{lambda_k}
\end{equation}
where $\lambda_{0}$ is the collapse rate. Throughout the text we will assume $\lambda_{0} = 1.03\times 10^{-5}\ \mathrm{Mpc}^{-1}$, which corresponds to a frequency scale $\tilde{\lambda_{0}} = c\,\lambda_{0} \simeq 10^{-19}\ \mathrm{s}^{-1}$ in MKS units. This choice ensures that $\lambda_{0}$–parameter lies within the region allowed by laboratory experimental constraints \cite{Ocampo_2024,PhysRevLett.124.080402}.
In turn, the extra added parameter $\alpha$ will account for the effects
from the CSL model. Moreover, recent works have pointed out that the additional
term $\alpha/c_{s}^{2}k^{2}$ induces similar estimates as a standard
$\Lambda$CDM model \citep{MICOL}.
\begin{figure}[t]
\includegraphics[scale=0.36]{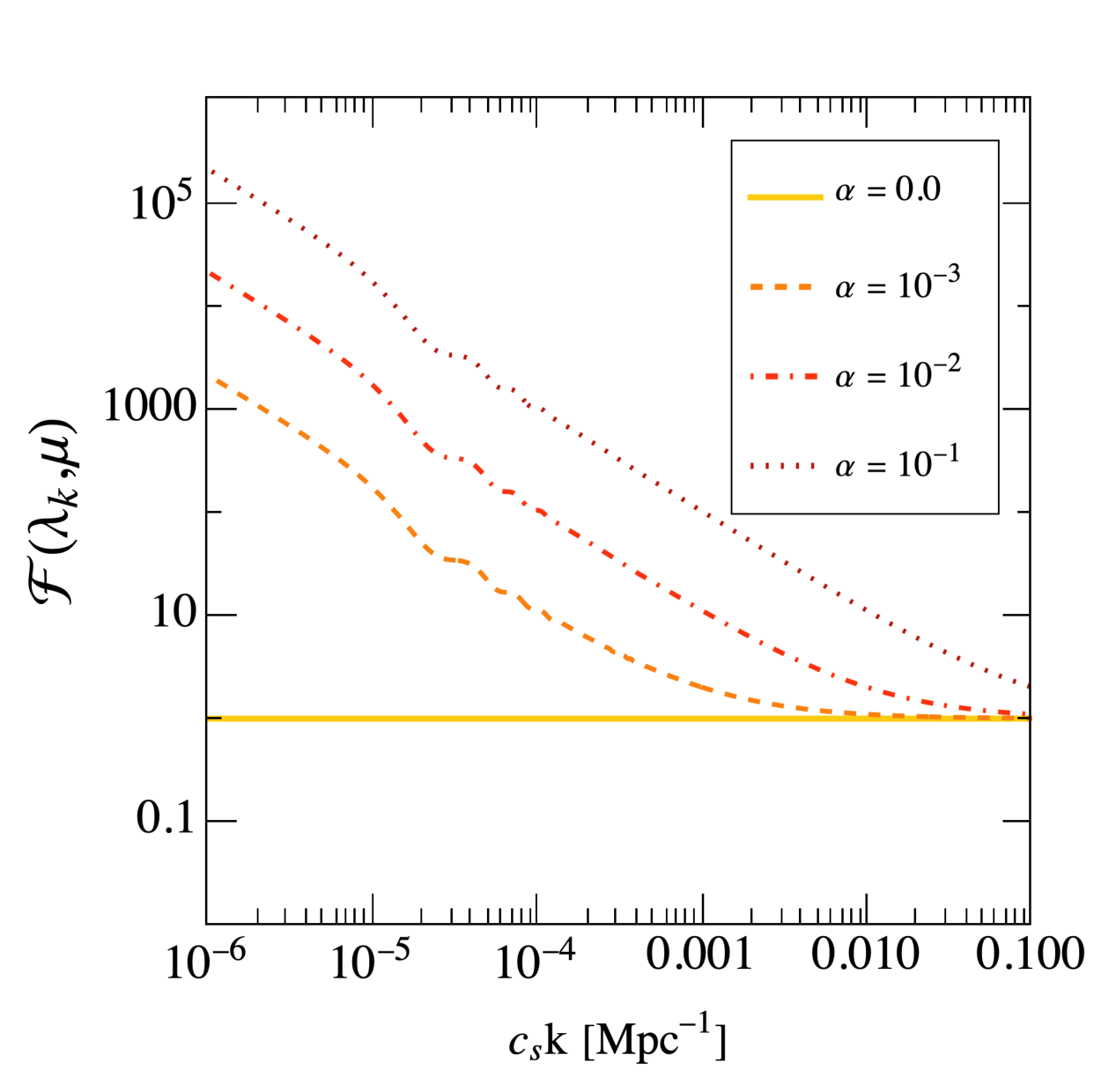} \caption{\label{fig3} In the context of the CSL inflationary model, the function
$\mathcal{F}(\lambda_{k},\nu_{s})$ is associated with the power spectrum.
The parameters $\lambda_{0}$ (set at $1/\left|\tau\right|=1.03\times10^{-5}\mathrm{Mpc^{-1}}$)
and $\mu$ (fixed at 0.5), remain constant. }
\end{figure}

\section{Power spectrum constraints from the CSL scheme}

In this section, we aim to analyze the deviations produced by the
collapse parameter from CSL theory on the scalar power spectrum. Thus,
by using the $\beta$-exponential potential (\ref{betaexponential}),
and (\ref{tn}) we can rewrite (\ref{scalarpwr}) as 
\begin{equation}
\mathcal{P}_{s}=\dfrac{\lambda^{2/(2\beta+1)}}{24}\left(\frac{V_{0}}{\pi^{2}M_{p}^{4}}\right)^{\frac{2\beta}{2\beta+1}}\left[1+2N(1-2\beta)\right]^{\frac{2\beta+2}{2\beta+1}}\mathcal{F}(\lambda_{k},\nu_{s}).\label{amplitude}
\end{equation}
Recent analysis, assuming that $\lambda_{k}$ must be positive, has
imposed constraints on $\alpha$-parameter such that $\alpha>-10^{-6}$
for the relevant $c_{s}k$ values \citep{PICCIRILLI}. Therefore,
replacing Eq. (\ref{lambda_k}) in (\ref{F_lambda_k}) -- for small
values of $\alpha$ -- one may now expand the collapse function in
first order on the $\alpha$-parameter, such that the function of
the collapse parameter becomes
\begin{equation}
\mathcal{F}(\lambda_{k},\mu)\simeq1+B(c_{s}k,\mu)\alpha,
\end{equation}
in which $\mu=\nu_{s},\nu_{t}$ and the function $B(c_{s}k,\mu)$
is given by 
\begin{eqnarray}
B(c_{s}k,\mu) & = & \frac{\lambda_{0}}{c_{s}^{2}k^{2}}\sin\gamma_{k}\cos\gamma_{k}-\frac{\lambda_{0}\tau}{2c_{s}k}\left(\dfrac{3}{\mu+1}\sin^{2}\gamma_{k}+\dfrac{\cos^{2}\gamma_{k}}{\mu}\right)\nonumber \\
 & - & \frac{4\mu\lambda_{0}^{2}}{c_{s}^{3}k^{3}(1+4\lambda_{0}^{2}/c_{s}^{2}k^{2})^{1+\mu/2}}\frac{\sin(\pi\mu)}{\sin(\mu\arctan(2\lambda_{0}/c_{s}k)-\pi\mu)}\nonumber \\
 & - & \frac{2\mu\lambda_{0}}{c_{s}^{2}k^{2}(1+4\lambda_{0}^{2}/c_{s}^{2}k^{2})^{1+\mu/2}}\frac{\sin(\pi\mu)\cos(\mu\arctan(2\lambda_{0}/c_{s}k)-\pi\mu)}{\sin^{2}(\mu\arctan(2\lambda_{0}/c_{s}k)-\pi\mu)}\mbox{.}
\end{eqnarray}

Noteworthy is that, for $\alpha=0$, then $\mathcal{F}(\lambda_{0},\nu_{s})=\mathcal{F}(\lambda_{0},\nu_{t})=1$,
and there is no modification on the standard shape of the power spectrum.
Moreover, the different values of $\alpha$ change the behavior of
$\mathcal{P}_{s}$ and $\mathcal{P}_{t}$, which implies modifications
in the spectral index of the inflation. In Fig. \ref{fig3} we present
the plot of the function of collapse parameter, $\mathcal{F}(\lambda_{k},\nu_{s})$,
for different values of $\alpha$. We note that at lower values of
$k$, the standard primordial power spectrum shape ($\alpha=0$) differs
significantly from the spectrum produced by the CSL collapse model.

In turn, the spectral index $n_{s}$ is defined by the relation
\begin{eqnarray}
n_{s}-1\equiv\frac{d\ln\mathcal{P}_{s}}{d\ln k}=\frac{d\ln\mathcal{A}_{s}}{d\ln k}+\frac{d\ln\mathcal{F}(\lambda_{k},\nu_{s})}{d\ln k}\,\mbox{,}
\end{eqnarray}
whereas, on the one hand, we have 
\begin{eqnarray}
\frac{d\ln\mathcal{A}_{s}}{d\ln k}=-2\varepsilon_{1}-\varepsilon_{2}\,\mbox{,}
\end{eqnarray}
and, on the other hand, the variation of the collapse function satisfies
\begin{eqnarray}
\frac{d\ln\mathcal{F}(\lambda_{k},\nu_{s})}{d\ln k}=-\frac{1}{1-\varepsilon_{1}}\frac{M_{p}^{2}V_{T}}{\alpha^{\prime}V^{2}}\frac{d\nu_{s}}{dT}\frac{d\ln\mathcal{F}(\lambda_{k},\nu_{s})}{d\nu_{s}}\,\mbox{.}
\end{eqnarray}
So, given that $\nu_{s}=1/2+\varepsilon_{1}+\varepsilon_{2}/2$, we
obtain 
\begin{equation}
\frac{M_{p}^{2}V_{T}}{\alpha^{\prime}V^{2}}\frac{d\nu_{s}}{dT}\simeq-\varepsilon_{1}\varepsilon_{2}-\varepsilon_{1}\varepsilon_{3}\mbox{,}
\end{equation}
while 
\begin{eqnarray}
\frac{d\ln\mathcal{F}(\lambda_{k},\nu_{s})}{d\nu_{s}}\simeq\alpha\frac{dB(c_{s}k,\nu_{s})}{d\nu_{s}}\approx\alpha\left[K_{0}+K_{1}f(\varepsilon_{1},\varepsilon_{2})\right]\,\mbox{,}
\end{eqnarray}
in which $K_{0}\equiv K_{0}(c_{s}k)$, $K_{1}\equiv K_{1}(c_{s}k)$,
and $\ensuremath{f(\varepsilon_{1},\varepsilon_{2})=\varepsilon_{1}+\varepsilon_{2}/2}$,
such that 
\begin{align}
K_{0}(c_{s}k) & =-\frac{\lambda_{0}\pi\left(2\sin(-2c_{s}k\tau)+\pi\cos(-2c_{s}k\tau)\right)}{4c_{s}^{2}k^{2}}+\frac{2\lambda_{0}\tau\left(\sin^{2}(-c_{s}k\tau)+3\cos^{2}(-c_{s}k\tau)\right)}{3c_{s}k}\nonumber \\
 & +\frac{\lambda_{0}\pi\ln B\left[c_{s}k+2\lambda_{0}(C-\pi)(\ln B-2)\right]}{c_{s}^{3}k^{3}B(C-\pi)^{2}},\\
K_{1}(c_{s}k) & =-\frac{\lambda_{0}\pi^{2}\tau}{c_{s}^{2}k^{2}}+\frac{8\lambda_{0}\tau}{9c_{s}k}\left(1+8\cos^{2}(-c_{s}k\tau)+3\pi\sin(-c_{s}k\tau)\cos(-c_{s}k\tau)\right)\nonumber \\
 & +\frac{4\pi\lambda_{0}^{2}\ln B}{c_{s}^{3}k^{3}B(C-\pi)}+\frac{\lambda_{0}\pi\left(4C^{2}-8C\pi-3\ln^{2}B+8\pi^{2}\right)}{6c_{s}^{2}k^{2}B(C-\pi)^{2}},
\end{align}
where $B=1+4\lambda_{0}^{2}/c_{s}^{2}k^{2}$, and $C=\arctan\left(2\lambda_{0}/c_{s}k\right)$.
The functions $K_{i}$ ($i=0,1$) are obtained by expanding the first
order derivative on the slow-roll parameters. Therefore, considering
an approximation to second order in the slow-roll parameters, we obtain
\begin{equation}
n_{s}=1-2\varepsilon_{1}-\varepsilon_{2}+K_{0}(\varepsilon_{1}\varepsilon_{2}+\varepsilon_{1}\varepsilon_{3})\alpha,\label{nsmod}
\end{equation}
where, for a pivot scale value $c_{s}k=0.05\,\mbox{Mpc}^{-1}$, $\tau=-9.71\times10^{4}\mbox{ Mpc}$,
and $\lambda_{0}=1.03\times10^{-5}\mbox{ Mpc}^{-1}$, we get $K_{0}=-31.30\mbox{ Mpc}^{-1}$.

In turn, for the ratio of tensor-to-scalar fluctuations, it becomes
\begin{eqnarray}
r\equiv\frac{\mathcal{P}_{t}}{\mathcal{P}_{s}}=\frac{\mathcal{A}_{t}}{\mathcal{A}_{s}}\frac{\mathcal{F}(\lambda_{k},\nu_{t})}{\mathcal{F}(\lambda_{k},\nu_{s})}\mbox{.}
\end{eqnarray}
Using the definitions $\nu_{s}=1/2+\varepsilon_{1}+\varepsilon_{2}/2$
and $\nu_{t}=1/2+\varepsilon_{1}$ and approximating in first order
on slow-roll parameters, we obtain 
\begin{eqnarray}
r=16\varepsilon_{1}\left(1+\alpha K_{2}\varepsilon_{2}\right)\mbox{,}\label{rmod}
\end{eqnarray}
with 
\begin{align}
K_{2}(c_{s}k) & =-\frac{\lambda_{0}\pi}{4c_{s}^{2}k^{2}}\sin(-2c_{s}k\tau)-\frac{2\lambda_{0}\tau}{3c_{s}k}\left(2+\cos(-2c_{s}k\tau)\right)+\frac{2\pi\lambda_{0}^{2}}{c_{s}^{3}k^{3}B(C-\pi)}\nonumber \\
 & -\frac{\lambda_{0}\pi\ln B}{2c_{s}^{2}k^{2}B(C-\pi)^{2}},
\end{align}
where $K_{2}=\left.\left(B(c_{s}k,\nu_{t})-B(c_{s}k,\nu_{s})\right)\right|_{c_{s}k=0.05}=31.28\mbox{ Mpc}^{-1}$,
for $\tau=-9.71\times10^{4}\mbox{ Mpc}$ and $\lambda_{0}=1.03\times10^{-5}\mbox{Mpc}^{-1}$.
Furthermore, we can verify that when turning off the CSL collapse
model effects, i.e., we take $\alpha=0$, relations (\ref{nsmod})
and (\ref{rmod}) recover the standard value, as expected.

Finally, with these results, by considering the $\beta$-exponential
potential, we get the scalar spectral index and tensor-to-scalar ratio,
as follows 
\begin{eqnarray}
n_{s} & = & 1-\frac{4(1+\beta)}{1+2N(1+2\beta)}+\frac{6K_{0}(1+2\beta)}{[1+2N(1+2\beta)]^{2}}\alpha,\label{indexesalpha}\\
r & = & \frac{16}{1+2N(1+2\beta)}+\frac{2K_{2}(1+2\beta)}{[1+2N(1+2\beta)]^{2}}\alpha.\label{r-eq}
\end{eqnarray}

\begin{figure}[t]
\includegraphics[scale=0.36]{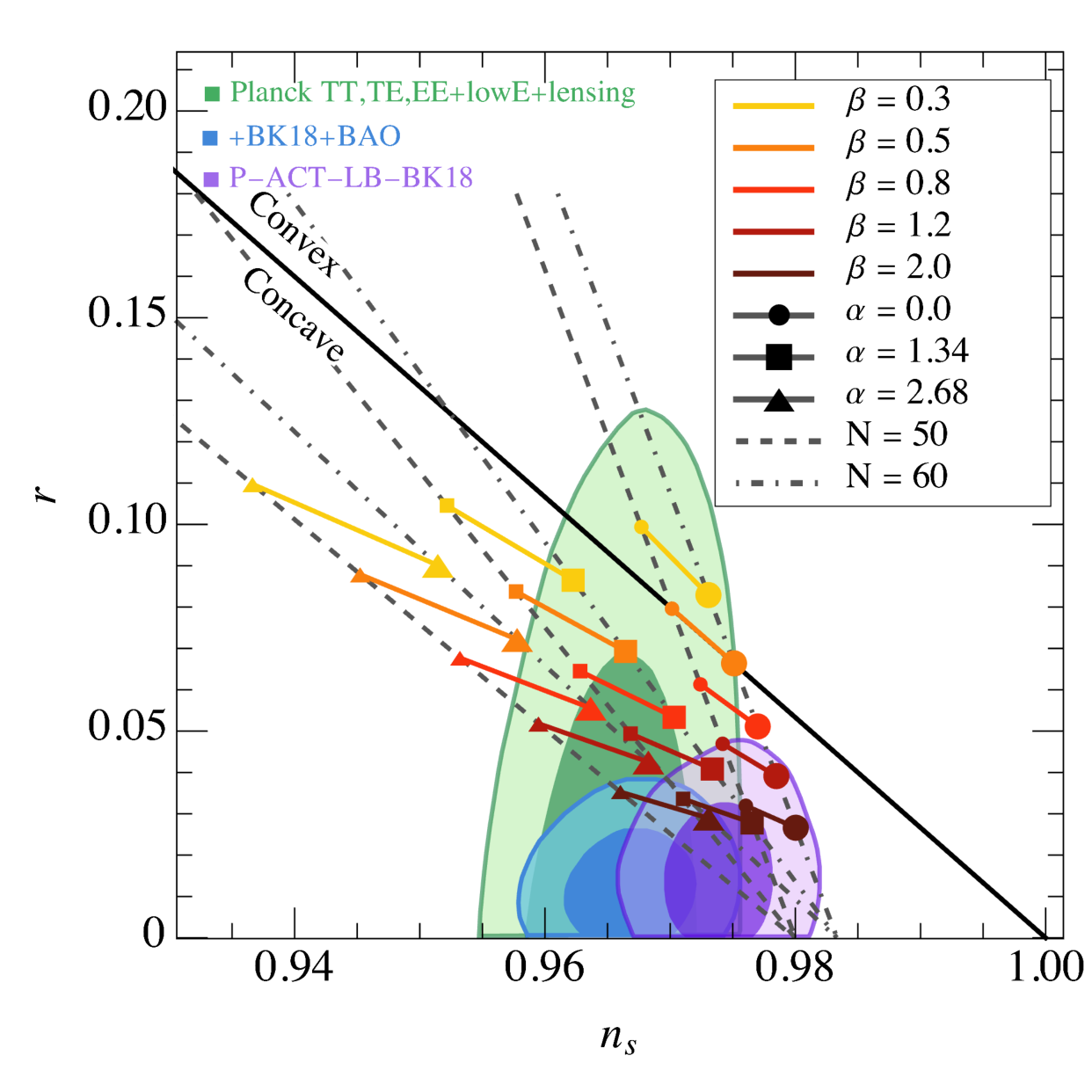}\caption{\label{fig4} The $n_{s}-r$ plane for the $\beta$-exponential potential
is shown for different values of $\alpha$ and $\beta$. The standard
tachyon inflation is recovered by assuming $\alpha=0$, while $\alpha\protect\neq0$
introduces the CSL inflationary model deviations, which changes the
behavior of the standard spectral index. Two values for the number
of e-folds, $N=50$ (dashed lines) and $N=60$ (dash-dotted curves),
are considered.}
\end{figure}

The deviations owing to the collapse parameter are shown in the $n_{s}-r$
plane (Fig. \ref{fig4}). The constraints shown in Fig. \ref{fig4}
for the Planck 2018 baseline analysis, adapted from \citep{PLANCK2018},
incorporating BICEP/Keck along with BAO data, present bounds that
exclude the tachyonic $\beta$-exponential inflation in the CSL approach
on determined ranges of $\alpha$ and $\beta$ parameters. In this
case, the contours in the vertical ($r$) direction are shrunk by
the BK18 data, whereas the BAO data shrinks the contours along the
horizontal ($n_{s}$) direction \citep{PLANCK2018}. At $95\%$ CL,
from Planck TT,TE,EE+lowE+lensing likelihood, for $N\geq50$, the results
impose bounds on the collapse parameter such that $\alpha<5.6\,\mathrm{Mpc^{-1}}$,
for $\beta=2.0$. Furthermore, from this analysis, one can observe
that, for $N\leq60$, we must have $\alpha<9.6\,\mathrm{Mpc^{-1}}$
with $\beta=2.0$. It turns out that the limits on $\alpha$ are strengthened by considering the ACT dataset. By assuming $\beta = 2.0$, so $\alpha < 2.7 \,(5.5)\,\mathrm{Mpc^{-1}}$ for $N\geq50\,(N\leq60)$.

On the other hand, the constraint for Planck measurements (BK15) \citep{PLANCK}
indicates that, for $N=50$, we have $\alpha=0.673\,\mbox{ Mpc}^{-1}$
and $\beta=0.635$, from the spectral index (\ref{indexesalpha})
and the tensor-to-scalar ratio (\ref{r-eq}), for $n_{s}=0.9658$
and $r=0.072$. In its turn, this analysis also allows us to impose
bounds on the $\beta$-parameter as long as we keep the collapse parameter
fixed. Indeed, in the case in which $\alpha=1.34\,\mathrm{ Mpc^{-1}}$,
for $N\leq60$, so $\beta\lesssim1.7$, while for $\alpha=2.68\,\mathrm{Mpc^{-1}}$,
we find that $\beta\lesssim2.8$ (for $N\leq60$). These constraints are relaxed if we consider bounds originating from ACT data \citep{calabrese2025atacamacosmologytelescopedr6}. In this case, one find $\beta\lesssim11.4$ for $\alpha = 1.34\,\mathrm{Mpc^{-1}}$, whereas $\alpha = 2.68\,\mathrm{Mpc^{-1}}$ imposes that $\beta\lesssim17.5$, both setting $N\leq60$. These results show
the clear relation between the $\alpha$–parameter and the other free
parameters of the $\beta$-exponential potential. As we saw, the CSL
approach weakens the constraints on the $\beta$-exponential model,
although it provides an excellent agreement between the theoretical
estimate and the current observational data. Thus, we note that while higher values of $\beta$ are favored by the constraints from ACT, higher values for the $\alpha$–parameter favor the limits imposed by Planck. Finally, it is worth
mentioning that even by adding the BK18 \& BAO data to the Planck
Collaboration $2018$ baseline analysis, or when considering the recent ACT data, one can not rule out the inflation
tachyon $\beta$–exponential model in the CSL scheme for a broad range of the parameter space.

\section{Concluding Remarks}

In this work, we have analyzed phenomenologically the constraints
on the collapse parameter, $\alpha$, by considering the tachyonic
inflationary $\beta$-exponential model in the context of the CSL
approach. Initially, we obtained constrained values for the standard
$\beta$-exponential model so that for $N\geq50$ ($N\leq60$), we
must have $\beta\leq2.0$ ($\beta\geq2\times10^{-2}$). By quantizing the scalar perturbations via the Mukhanov-Sasaki formalism and applying the CSL model, we have obtained
an important modification in both the amplitude and shape of the primordial
power spectrum, which presents dependence on the strength of the collapse
parameter, $\alpha$. We show that if one turns off the quantum effects
owing to the collapse parameter, i.e., when $\alpha=0$, one recovers
the expected result from tachyonic inflation.

Finally, the modifications owing to the CSL scheme, when applied to
the $\beta$-exponential potential, lead to deviations in the spectral
indexes, which, in principle, could yield a clear fingerprint of tachyonic
inflation on the CMB spectrum. From this analysis, we were able to
impose bounds on the collapse parameter. Indeed, for a realistic number
of e-folds before the end of inflation, i.e., $N \geq 50\, (N \leq 60)$,
and for $\beta=2.0$, one obtains that $\alpha<5.6\,\mathrm{Mpc^{-1}}\left(9.6\,\mathrm{Mpc^{-1}}\right)$ from Planck TT,TE,EE+lowE+lensing dataset, whereas the recent ACT result provides us $\alpha<2.7\,\mathrm{Mpc^{-1}}\left(5.5\,\mathrm{Mpc^{-1}}\right)$. It turns out that, by considering the CSL approach, we have shown
that the constraints on the $\beta$-parameter become weak. In this case,
if $\alpha=1.34\,(2.68)\,\mathrm{Mpc^{-1}}$ $\left(N\leq60\right)$,
so one gets $\beta\lesssim 1.7\, (2.8)$. The inclusion of ACT data notably expands the allowed parameter space. Specifically, for $N \leq 60$, we find that for a collapse parameter of $\alpha = 1.34\,\mathrm{Mpc^{-1}}$, values up to $\beta \lesssim 11.4$ are allowed, while for $\alpha = 2.68\,\mathrm{Mpc^{-1}}$, this limit is weakened to $\beta \lesssim 17.5$. As we see, this
study opens an avenue to investigate inflation from other tachyonic
potentials by considering the context of the CSL scheme, aiming to
obtain new constraints on the collapse parameter. 
\begin{acknowledgments}
We would like to thank CNPq, CAPES and CNPq/PRONEX/FAPESQ-PB (Grant
No. 165/2018), for partial financial support. FAB acknowledges support
from CNPq (Grant No. 309092/2022-1). JCMR acknowledges support from
CAPES. ASL acknowledges support from CAPES (Grant No. 88887.800922/2023-00).
ASP thanks the support of the Instituto Federal do Pará. 
\end{acknowledgments}

\bibliography{apssamp}

\end{document}